\documentclass[aps,prl,twocolumn,superscriptaddress,groupedaddress]{revtex4}

\usepackage{graphicx}  
\usepackage{amsmath}
\usepackage{bm}        
\usepackage{amssymb}   

\usepackage{float}

\usepackage{color}

\begin{document}

 \title{Fabricating large two-dimensional single colloidal crystals by doping with active particles}
 
\date{\today}
 \author{B. van der Meer}
 \author{L. Filion}
 \author{M. Dijkstra}

\affiliation{Soft Condensed Matter, Debye Institute for Nanomaterials Science, Utrecht University, Princetonplein 5, 3584 CC Utrecht, The Netherlands}

\begin{abstract}
 
Using simulations we explore the behaviour of two-dimensional colloidal (poly)crystals doped with active particles. We show that these active dopants can provide an elegant new route to removing grain boundaries in polycrystals. Specifically, we show that active dopants both generate and are attracted to defects, such as vacancies and interstitials, which leads to clustering of dopants at grain boundaries. The active particles both broaden and enhance the mobility of the grain boundaries, causing rapid coarsening of the crystal domains. The remaining defects recrystallize upon turning off the activity of the dopants, resulting in a large-scale single-domain crystal.

%
%
%


\end{abstract}
 
\maketitle

\section{Introduction}

Colloidal crystals are ordered, self-assembled solids comprised of (sub)micron spheres that are arranged on a periodic lattice. 
Such crystals are almost always polycrystalline in nature and feature grain boundaries that separate adjacent crystal domains of different crystallographic orientations~\cite{gokhale2013grain}. Unfortunately, the presence of these grain boundaries is well known to drastically alter the material properties compared to a single crystal, for instance altering the photonic properties of the crystal~\cite{ye2001self}.  Avoiding the formation of grain boundaries during the fabrication of colloidal crystals is  very challenging and typically relies on either controlled growth of a single domain ~\cite{cheng1999controlled,ye2001self},  epitaxial growth from a template~\cite{van1997template,lin2000entropically,braun2001epitaxial,allard2004colloidal}, or applying external fields~\cite{palberg1995grain,amos2000fabrication,gokhale2012directional}.  Hence, new methods to remove grain boundaries are highly desirable. Here, we examine an elegant, new avenue for removing grain boundaries: doping colloidal polycrystals with self-propelled particles. 

Self-propelled particles, also known as active particles, incessantly convert energy into self-propulsion, and as such are intrinsically out-of-equilibrium. While traditionally such particles occured solely within the purview of natural systems (e.g. bacteria~\cite{sokolov2012physical,schwarz2012phase}), recent experimental breakthroughs have led to many novel types of synthetic colloidal swimmers and self-propelled particles~\cite{paxton2004catalytic,dreyfus2005microscopic,howse2007self,theurkauff2012dynamic,buttinoni2013dynamical,palacci2013living}.
These systems exhibit a wealth of new phase behaviour, including, for example, motility-induced phase separation into dense and dilute phases~\cite{theurkauff2012dynamic,palacci2013living,buttinoni2013dynamical,stenhammar2013continuum,stenhammar2014phase,wittkowski2014scalar,bialke2013microscopic,redner2013structure,speck2014effective,fily2012athermal,fily2014freezing},  giant density fluctuations \cite{narayan2007long,deseigne2010collective}, and swarming \cite{schaller2010polar}.
 Moreover, experimental and simulation studies have shown that the dynamics of a passive system can be altered dramatically by incorporating as little as $1\%$ of active particles into the system~\cite{ni2014crystallizing,kummel2015formation}.  At these concentrations, the self-propelled particles can be viewed as active ``dopants", which like passive dopants, can strongly alter the properties (e.g. dynamics) of the underlying passive system. Specifically, active dopants have been shown to promote  crystallization in hard sphere glasses~\cite{ni2014crystallizing} and it has recently been suggested that active dopants can be used to help fabricate large, defect-free crystals~\cite{kummel2015formation}. Here we confirm this possibility using computer simulations of dense mixtures of active and passive spherical colloids, and explore the mechanism responsible for the growth of crystal domains.

 
 
 \section{Method and Model}
 
 \begin{figure*}[!hbt] 
\includegraphics[width=1.0\textwidth]{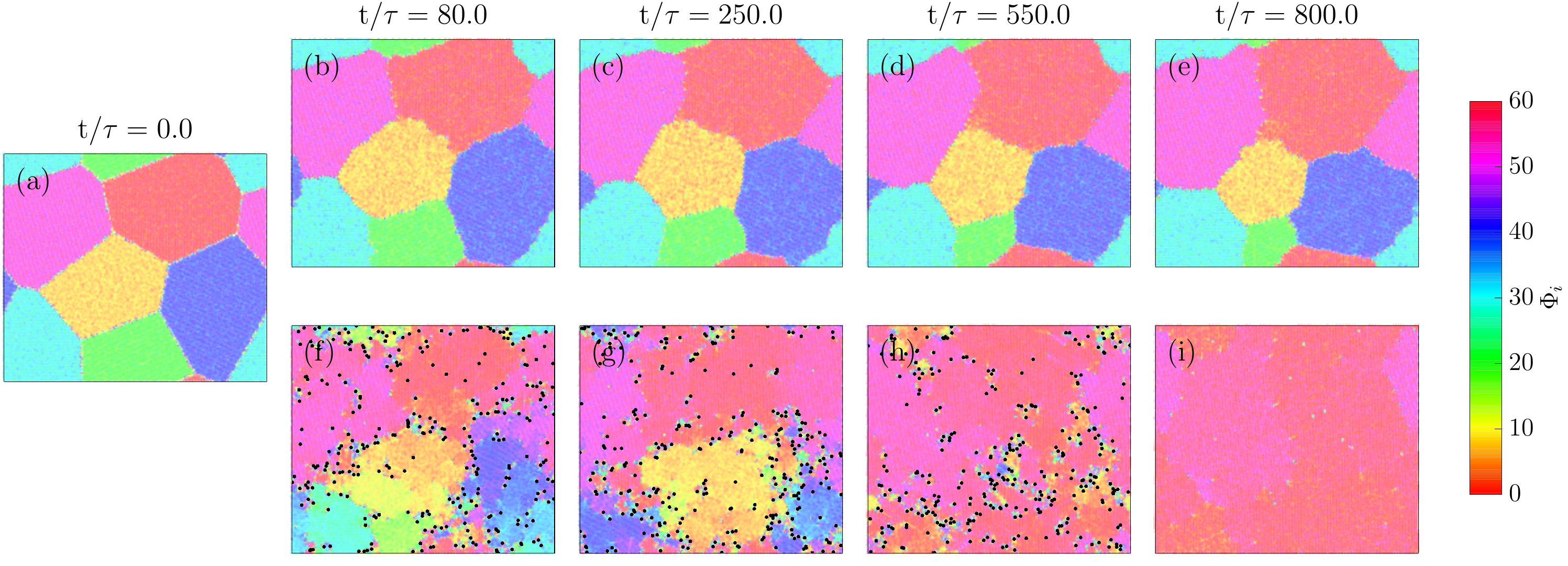} 
\caption{Grain growth in a passive system (a-e) vs a doped system (a,f-i) starting from initial configuration (a), showing significant coarsening of crystal domains in the doped system, compared to its purely passive counterpart. Here the color indicates the local orientation of a particle.  In the lower panel we activate a fraction $\alpha=0.0076$ of  particles for $0<t/\tau<550$ with a self-propulsion of $f\sigma /k_BT=90$. Active particles are plotted in black at many times their diameter to improve their visibility. After the activity of the self-propelled particles is switched off we can clearly see that the polycrystal has coarsened. System size: $N=39446$ and the density is  $\rho \sigma^2= 0.842$.}
\label{GBs}
\end{figure*}

\begin{figure*}[hbt!] 
\includegraphics[width=0.93\textwidth]{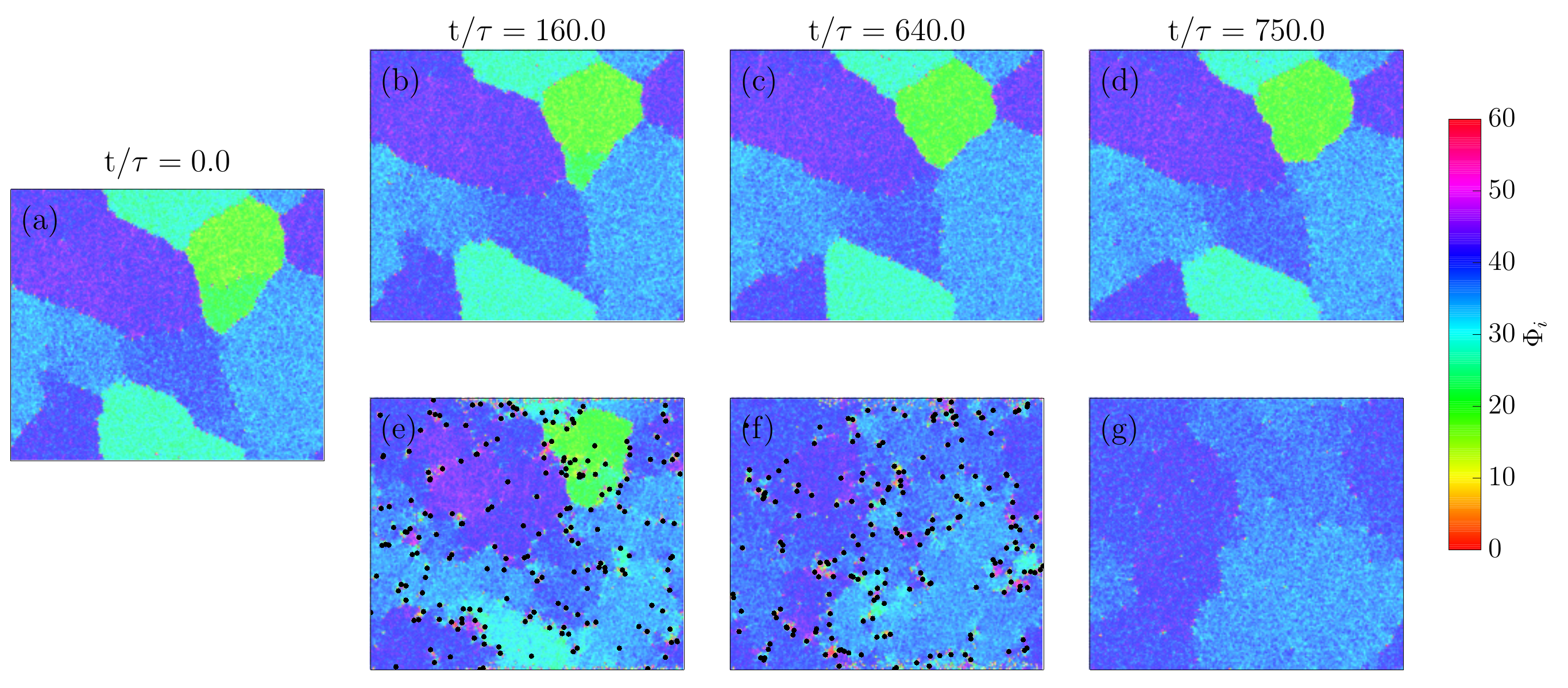} 
\caption{Grain growth in a system doped with active particles (b-d) vs a purely passive system (e-g) starting from initial configuration (a), showing significant coarsening of crystal domains in the doped system, compared to its purely passive counterpart.  In the lower panel we activate a fraction $\alpha=0.00264$ of particles for $0<t/\tau<640$. Active particles are plotted in black at many times their diameter to improve their visibility. After the activity of the self-propelled particles is switched off for a while we can clearly see that the polycrystal has been coarsened. System size: $N=75625$ and the density is $\rho \sigma^ 2=0.825$.}
\label{GBq}
\end{figure*}

We perform Brownian dynamics simulations of two-dimensional crystals consisting of $N$ particles interacting with a purely repulsive Weeks-Chandler-Andersen (WCA) potential 
\begin{equation}
  \beta U_{WCA}(r)= \left\{ 
  \begin{array}{ll}
  4\beta \epsilon \left[(\frac{\sigma}{r})^{12}-(\frac{\sigma}{r})^{6}+\frac{1}{4} \right], & r /\sigma \leq 2^{1/6} \\
  0, &  r /\sigma > 2^{1/6} 
  \end{array}
  \right.
\end{equation}
with $\sigma$ the particle diameter, $\beta \epsilon=40$ the energy scale, and $\beta=1/k_BT$, where $k_B$ is the Boltzmann constant and $T$ is the temperature. We activate a small fraction of randomly selected particles $\alpha = N_{a} /N$ by adding a constant self-propulsion force $f$ along the self-propulsion axis $\hat{\textbf{u}}_i$. The direction of this axis $\hat{\textbf{u}}_i$ rotates with the Brownian rotational diffusion coefficient $D_r = 3D_0 /\sigma^2$ with $D_0$ the short-time diffusion coefficient. For passive particles $f  = 0$.  We measure time in units of the short-time diffusion $\tau=\sigma^2/D_0$.


\section{Results}
\begin{figure*}
\includegraphics[width=\textwidth]{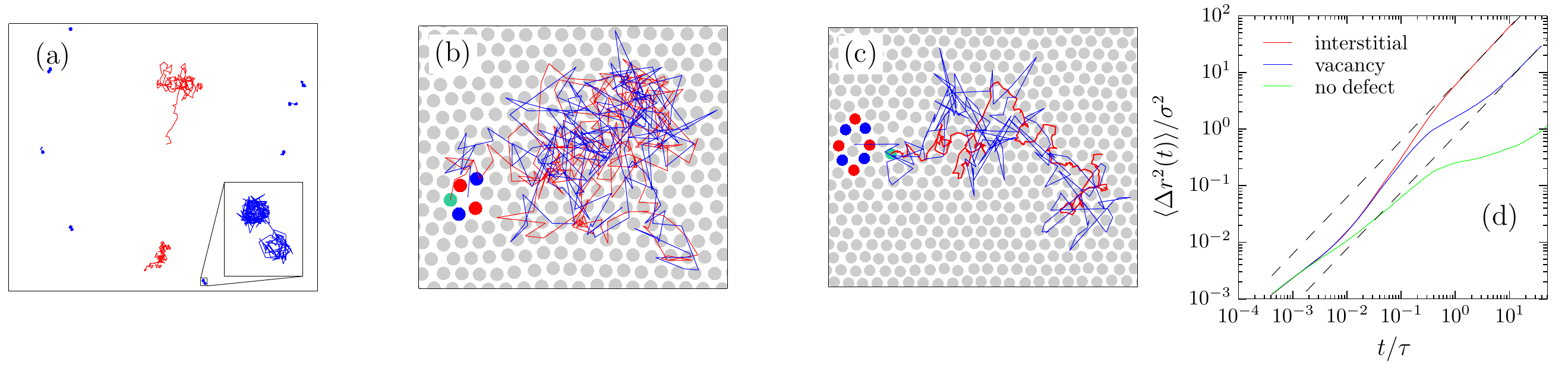} 
\caption{ (a) Trajectories of active particles with $f\sigma/k_BT=50$ at $\rho \sigma^2=0.825$ integrated over $t/\tau=100$. Red and blue trajectories indicate mobile and caged particles, respectively. (b,c) The trajectories of mobile self-propelled particles that are carrying a vacancy (b) or an interstitial (c) for $f\sigma/k_BT=50$ at $\rho \sigma^2=0.825$. In the configurations in (b) and (c) the active particle is highlighted in green. Defected particles are shown in red and blue, for seven-fold and five-fold coordination, respectively. (b) The active particle trajectory (red) and vacancy trajectory (blue) integrated over $t/\tau$=606.  (c) The active particle trajectory (red) and interstitial trajectory (blue) integrated over $t/\tau$=45.6. (d) Mean square displacement $\langle \Delta r^2(t) \rangle$ for a self-propelled particle $f \sigma/k_BT=50$ with a vacancy, an interstitial, and no point defect at $\rho \sigma^2=0.825$.  }
\label{vitrajs}
\end{figure*}

We begin by examining the effect of active dopants on grain boundaries. 
In Figure \ref{GBs} we show a time series of the evolution of a polycrystal in thermal equilibrium ($\alpha=0$) and in the case where some particles are made active ($\alpha = 0.0076$). The initial polycrystal, shown in Figure \ref{GBs}(a), was created by filling polygonal domains with crystallites of different orientations. In the snapshots in Figure \ref{GBs}, we color code particles according to their local orientation using $$\Phi_{i}=\dfrac{1}{\mathcal{N}_i} \sum \limits_{j=1}^{\mathcal{N}_i} \theta_{ij},$$ with $\mathcal{N}_i$ the number of nearest neighbours of particle $i$ as determined by a Voronoi construction, and the angle $\theta_{ij}=\tan^ {-1}({\bf r}_j-{\bf r}_i)$ with ${\bf r}_i$ the position of particle $i$ and $-\pi\leq \theta_{ij} \leq \pi$ ~\cite{skinner2010grain}.  In Figure \ref{GBs}(a-e) we show how the passive system evolves in time. Clearly, in thermal equilibrium ($\alpha = 0$)  the grain structure evolves very slowly: the thermal fluctuations allow for some motion of the grain boundaries over long periods of time, yet their mobility is very low, and as such little grain growth occurs. However, as shown in Figures \ref{GBs}(f-h), upon activation of a small number of particles, $\alpha=0.0076$, we observe a large increase in the grain boundary mobility, leading to significant coarsening of the domains over time. Interestingly, whereas in thermal equilibrium the grains are separated by a thin,  smooth interface, the grain boundaries in the activated systems are highly fluctuating and much broader, reminiscent of the local melting of the grain surfaces observed in  Ref. ~\onlinecite{kummel2015formation}. Note that we also examined the evolution of polycrystals that were obtained through crystallization of a supercooled fluid (instead of the artificially formed polycrystals as shown in Figure \ref{GBs}) and observed no significant differences. The time evolution of such a system is shown in Figure \ref{GBq}.

From Figure \ref{GBs} and \ref{GBq} it is clear that upon activation, self-propelled particles accumulate at the grain boundaries.  This accumulation of self-propelled particles enables the rearrangement of many particles at the surface of the grains leading to significant  motion of the boundary over time. In the process, the grain boundaries essentially disappear, and what remains is a single-domain crystal which now contains many local defects as the active particles are constantly perturbing the crystal lattice. The active particles have thus restructured the solid and annealed out the otherwise persistent grain boundaries, at the expense of creating other local defects.

Importantly, by switching the activity off, i.e. making all particles passive again ($f=0$) [Figure \ref{GBs}(i)], we can now obtain an almost defect-free single crystal. Experimentally one may switch off the active particles by using particles that are responsive to certain stimuli (e.g. light-activated colloids) or simply by letting all the fuel be consumed.


An important question that arises in looking at Figures \ref{GBs} and \ref{GBq} is how active particles that are located within a domain migrate to the grain boundaries to mediate the active grain growth.
Hence, to better understand the dynamics of the active dopants within the grains, we now focus on the behaviour of active particles in a single crystal. 
In Figure \ref{vitrajs}(a), we plot the trajectories of active particles in such a single crystal domain. 
We clearly observe two distinct types of trajectories: some active particles migrate large distances over extended periods of time (red trajectories), while other active particles remain caged by their passive neighbours (blue trajectories).

To disentangle this mobile/immobile duality we follow the topological structure of the solid over time using a Voronoi construction which allows us to locate, in addition to the active particles, point defects in the crystals. From studying the topological structure of various snapshots, we observe that the immobile self-propelled particles are trapped inside defect-free hexagonal cages while the mobile self-propelled particles migrate by carrying a vacancy or interstitial. In Figure \ref{vitrajs}(b,c) we plot the trajectories of a single active particle interacting with a single interstitial and vacancy, respectively. Note that in the configurations displayed in Figure \ref{vitrajs}(b,c) the active particle is highlighted in green and particles with seven-fold and five-fold coordination are shown in red and blue, respectively. Clearly, the active particle and the defects remain in close proximity to each other.  The strong overlap between the trajectories suggest that the defect and the active particle are attracted to each other. 

Note that these defects form spontaneously due to the activity  of the  self-propelled particles, even when we start from a perfect single crystal. These defects diffuse along with the active particle as it moves through the crystal, allowing it to remain mobile for long periods of time.  Hence, our data illustrate that active particles not only interact very strongly with the grain boundaries, but also with many other types of defects within the crystal such as (multi) vacancies and interstitials. These defects thus play a pivotal role in determining the mobility of the active particles.

\begin{figure*} 
\includegraphics[width=0.9\textwidth]{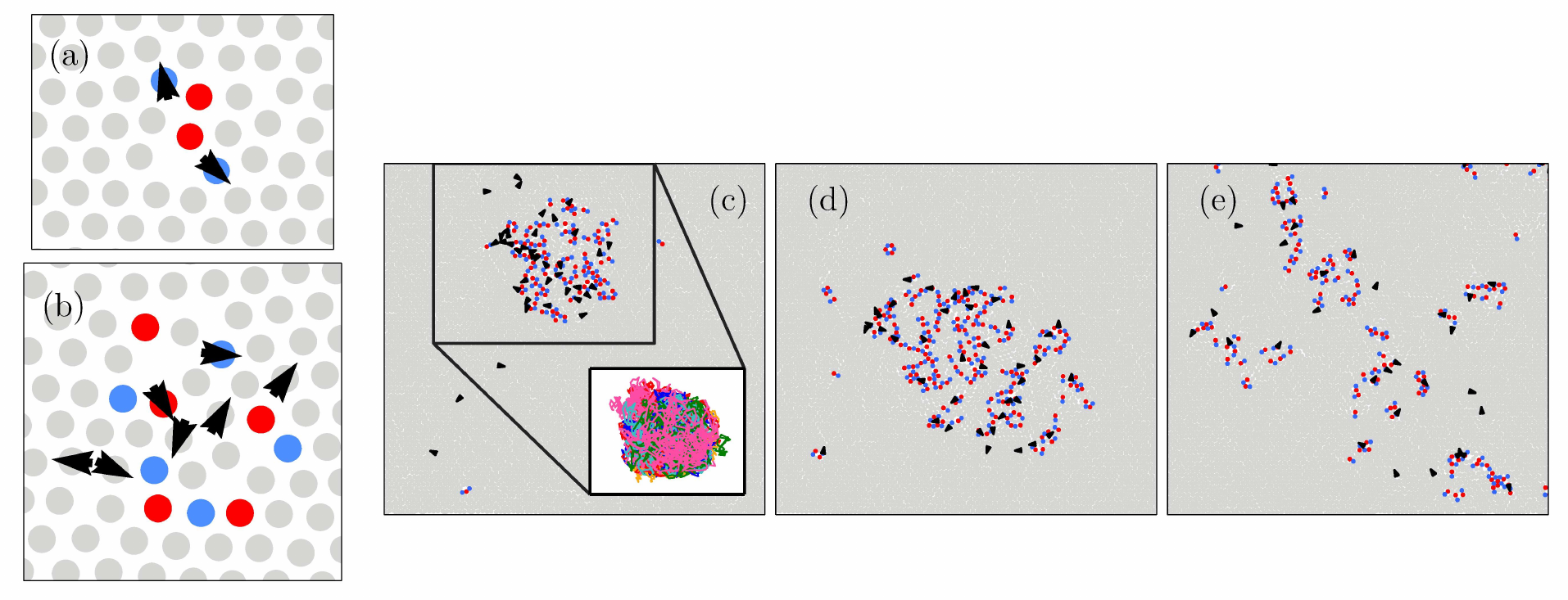} 
\caption{Snapshots of the system for increasing activity. We color-code non-sixfold coordinated particles with five or less nearest neighbours as blue and particles with  seven or more neighbours as red. The black arrows are superimposed on the active particles. (a) Two active particles sharing a single vacancy. (b) Seven active particles at one end of a defect string. (c) At sufficient amounts of active particles $\alpha=0.004$  a single, large defect cluster forms that is rich in active particles. (d) Upon increase of the self-propulsion $f \sigma/k_BT=70$  the defect cluster starts to grow. (e) At very high activity $f \sigma/k_BT=90$ defects and active particles are found throughout the crystal.  Note that the density $\rho \sigma^2=0.845$ for (a-e) and the fraction of active particles $\alpha=0.002$ for (a,b) and $\alpha=0.004$ for (c-e). Particles are shown slightly smaller or bigger for visual purposes.}
\label{singlecrystalsystem}
\end{figure*} 

To quantify the increase in mobility of self-propelled dopants upon picking up point defects, we consider the mean square displacement $\langle \Delta r^2(t) \rangle= \langle | \textbf{r}(t) - \textbf{r}(0) |^2 \rangle$ of a self-propelled particle carrying a vacancy, an interstitial, and no point defect. Self-propelled particles with no defect are caged resulting in a clear plateau in $\langle \Delta r^2(t) \rangle$, while self-propelled particles with a defect exhibit substantial diffusion [Figure \ref{vitrajs}(d)]. These self-propelled particles with point defects exhibit an intermediate super-diffusive regime, scaling as $\langle \Delta r^2(t) \rangle \sim t^b$ with $1<b\lesssim1.5$. Such a super-diffusive regime originates from the ballistic motion inside the cage or during cage jumping. While self-propelled particles with interstitials enter a diffusive regime soon after hopping, when $\langle \Delta r^2(t) \rangle \approx \sigma$, self-propelled particles with vacancies exhibit sub-diffusive motion after having travelled one lattice spacing. This feature arises from the fact that self-propelled particles with a vacancy become trapped inside the vacancy for a while. As a consequence, self-propelled particles have an enhanced probability of rotating their axis of self-propulsion and jumping back into the vacancy, thus leading to a sub-diffusive regime. Eventually, the diffusion mediated by the vacancies also becomes Fickian, scaling as $\langle \Delta r^2(t) \rangle \sim t$. 

The interactions between active particles and crystal defects can also generate attractive interactions between the active particles themselves. For instance, in Figure \ref{singlecrystalsystem}(a,b),  we show a few snapshots where active particles share defects for $f\sigma/k_BT=50$ and $\alpha=0.002$. Here, the active particles are marked with superimposed arrows that point in the direction of their self-propulsion axis. We observe that single vacancies or interstitials are often shared between two active particles, as shown in Figure \ref{singlecrystalsystem}(a) leading to attractions.


The attractive interactions between point defects in all combinations (vacancy-vacancy, interstitial-interstitial, and vacancy-interstitial) can lead to the formation of linear defect strings~\cite{lechner2009defect,lechner2013self}. We find that active particles typically  accumulate at the end of these defect strings and we show one such example in Figure \ref{singlecrystalsystem}(b). Here many additional defects have formed a small cluster at the end of the string, owing to the fact that the active particles are constantly exciting the end of the defect string.  
If we increase the fraction of active particles $\alpha=0.004$ we observe that these defect clusters grow and eventually coalesce to form one large defect cluster that is rich in active particles, as shown in Figure \ref{singlecrystalsystem}(c). The attractive interactions between point defects thus drives the system to "phase separate" into a hexagonal crystal and a complex defect cluster which is rich in active particles. We plot the trajectories of some active particles that are inside the big defect cluster in the inset of Figure \ref{singlecrystalsystem}(c). Clearly, the active particles become trapped inside the defect cluster and are unlikely to leave the defected area. Nonetheless, they can move throughout the whole defect zone, migrating from one side of the cluster to the other.  The size of the defect zone increases with  increasing self-propulsion [Figure \ref{singlecrystalsystem}(d)]. However, upon increasing the self-propulsion $f$ further this cluster breaks up, as shown in Figure \ref{singlecrystalsystem}(e).



\begin{figure}
 \begin{tabular}{lcl}
 \includegraphics[width=0.43\linewidth]{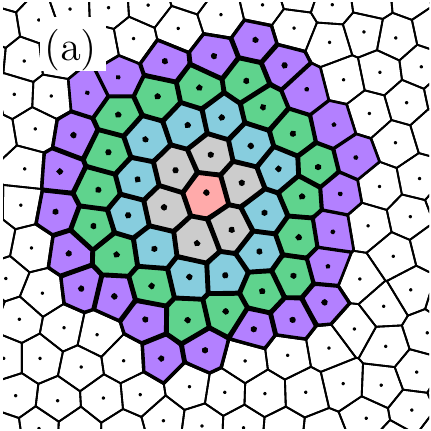} & &
 \includegraphics[width=0.40\linewidth]{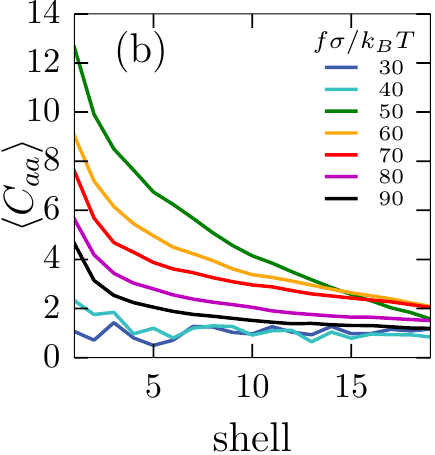}
 \end{tabular}
 
 \caption{ (a) Concentric shells of Voronoi cells around a central particle. Here particles in the same shell have the same minimal path length within the Voronoi construction. (b) Normalized fraction of active neighbours of an active particle in concentric Voronoi shells $\langle C_{aa} \rangle $ for different amounts of self-propulsion $f \sigma/k_BT$.}
\label{topo}
  
\end{figure}

To further explore the attraction between active particles, we determine the fraction of active neighbours of an active particle in concentric Voronoi shells. Specifically, for each active particle $i$, we assign each other particle in the neighbourhood of particle $i$ to a shell $s$, based on the shortest path to particle $i$ within the Voronoi construction (see  Figure \ref{topo}(a)).  We then calculate the fraction of active particles within each shell, normalized by the remaining fraction of active particles in the system,  i.e. $C_{aa}(s)= g_{aa}(s) / g_{ideal},$ where $g_{aa}(s)$ is the average fraction of active particles in shell $s$ around an active particle, and $g_{ideal}=\dfrac{N_a -1}{N-1}$ which in the thermodynamic limit equals $\alpha$. We plot $C_{aa}(s)$ for various self-propulsion forces $f$ in Figure \ref{topo}(b).  For low self-propulsion $f \sigma/k_BT \leq 30$ the active particles are distributed (almost) ideally, i.e. $C_{aa}\approx 1$ for all $s$. At  $f \sigma/k_BT= 50$ we observe a large peak in $C_{aa}$ at small $s$, showing that active particles are preferentially found in close proximity: active particles cluster, in agreement with Figure \ref{singlecrystalsystem}(c). Upon further increasing the self-propulsion we find that the peak in $C_{aa}$ lowers and starts to decay more slowly. This corresponds to an expansion of the defect clusters which distributes the active particles over a larger area. At high self-propulsion $f \sigma/k_BT \geq 80$ we see that the clustering becomes much less pronounced, as can also be seen from Figure \ref{singlecrystalsystem}(e).

\begin{figure*} 
\includegraphics[width=1.0\textwidth]{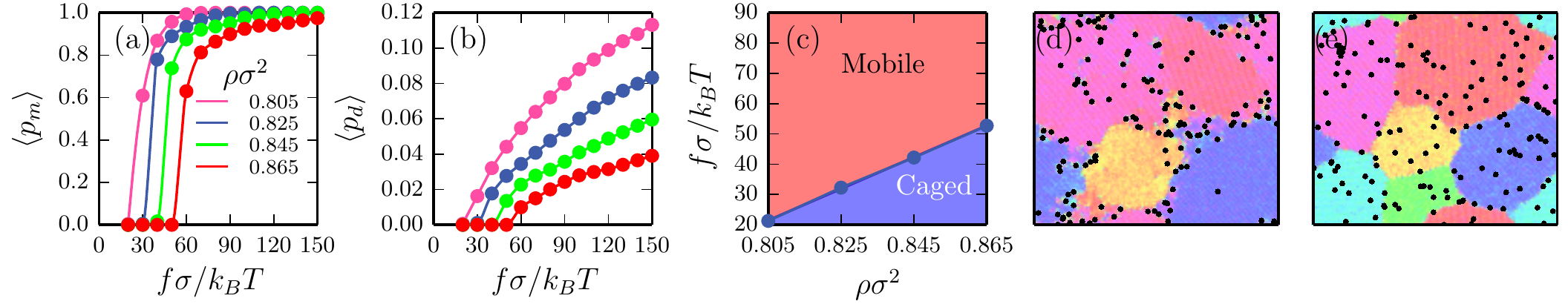} 
\caption{  (a) The average fraction of mobile self-propelled particles $\langle p_m \rangle$. (b) The average fraction of defects $\langle p_d \rangle$. (c) State diagram indicating the regimes where active particles are caged vs mobile.  (d) In the mobile regime, $\rho \sigma^2=0.845$ and $f \sigma /k_BT=90$,  we observe significant coarsening of polycrystals. Here active particles accumulate at the grain boundaries. (e) In the caged regime, $\rho \sigma^2=0.845$ and $f \sigma /k_BT=30$,  the grain growth occurs very similarly to the purely passive system. Here active particles remain inside the crystal domains and no enhanced grain growth occurs. Color-coding the same as in Figure \ref{GBs}. $\alpha=0.004$ in all plots.}
\label{mobility}
\end{figure*}
 
Intuitively, one may expect that at low self-propulsion all self-propelled particles are caged. On the other hand, if the self-propulsion is  high we expect active particles to be mobile. 
To examine the presence or absence of this  dynamical transition we calculate the fraction of mobile active particles $\langle p_m \rangle$, defined as particles that have moved more than one lattice spacing $\Delta r \ge a$ within a sufficiently long time interval $\Delta t=2.5\tau$ \footnote{Note that we checked that our results are robust to variations in both the displacement threshold as well as the time-interval.}. In Figure \ref{mobility}(a) we plot $\langle p_m \rangle$ as a function of $f$ for varying densities.  While at low self-propulsions $f$ all active particles remain caged $\langle p_m \rangle=0$, we observe a sudden increase in the average  number of mobile active particles beyond some critical $f$, indicating a sudden onset of cage breaking [Figure \ref{mobility}(a)]. With increasing density $\rho$ the lattice becomes more rigid, making the jump in $\langle p_m \rangle$ shift to higher $f$ as the active particles require more force in order to break from their cage. 

\begin{figure} 
\includegraphics[width=0.35\textwidth]{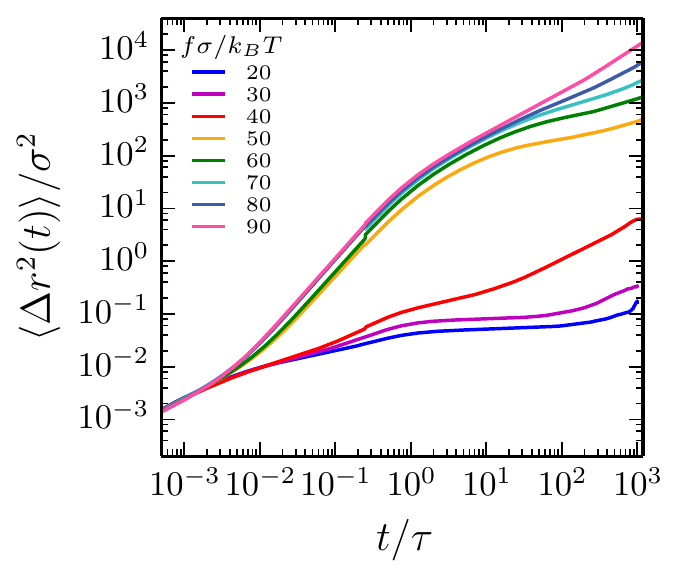} 
\caption{Mean square displacement $\langle \Delta r^2(t) \rangle$ of self-propelled particles at $\rho \sigma^2=0.845$ with an active fraction of $\alpha=0.004$ for different strengths of self-propulsion $f\sigma/k_BT$.}
\label{msdsysact}
\end{figure}

We further quantify the dynamics of self-propelled particles for various self-propulsions, we calculate the mean square displacement, see Figure \ref{msdsysact} for $\alpha=0.004$ at $\rho \sigma^2=0.845$. While, at very short time scales ($t/\tau \lesssim 10^{-3}$) all mean square displacements overlap, we can see that at longer time scales the dynamics of active particles are separated into two groups. Clearly, at low self-propulsions $f \sigma /k_BT\leq 40$ particles remain (more or less) caged, while at higher self-propulsion  the mean square displacement of the active particles exhibits a strong increase over time.  Such an increase of the long-term mean square displacement with increasing activity is in good agreement with the data shown in Figure \ref{mobility}(a).

We also observe that the increase in mobility of active particles strongly correlates with the formation of defects, as monitored by the average fraction of defects $\langle p_d \rangle$ inside the crystals, shown in Figure \ref{mobility}(b). The onset of defect formation ($\langle p_d \rangle >0$) coincides perfectly with the large jump observed in $\langle p_m \rangle$. With increasing density $\rho$, the crystal becomes stiffer and defects form at higher self-propulsions. The interplay between the self-propulsion $f$ and the density $\rho $, which respectively promote and restrict diffusion of active particles, is captured in a state diagram, as shown in Figure \ref{mobility}(c). Here we used the average fraction of mobile active particles as a cutoff $\langle p_m^* \rangle > 0.1$ to differentiate between the two regimes where active particles are caged or mobile. 

This cage-breaking transition plays a crucial role in active grain growth. Specifically we only observe significant active grain growth in systems in which the active particles are mobile. Upon activation, self-propelled particles accumulate at the grain boundaries, causing many particles at the surface of the grains to rearrange, thus leading to significant grain boundary motion over time [Figure \ref{mobility}(d)]. However, in systems where active particles are caged, the active particles cannot diffuse to the grain boundaries [Figure \ref{mobility}(e)] and we observe no such enhancement in the grain growth; the grain growth  is very similar to a purely passive system. Hence, active particles enhance grain growth only when they can easily diffuse to the grain boundaries.

\subsection{Conclusions}
 
In conclusion, we have explored the doping of two-dimensional colloidal (poly)crystals with sparse active particles.  We find that active particles enhance the mobility of grain boundaries, leading to large-scale single-domain crystals containing local defect clusters rich in active particles. Further analysis on single crystals revealed an intricate relation between the dynamics of active particles and the formation, and motion of defects in two-dimensional colloidal crystals. Not only do defects play a pivotal role in determining the mobility of active particles, they also generate attractions between active particles, which eventually drive the system to ``phase separate'' into mostly passive hexagonal crystal domains and complex defect clusters that are rich in active particles. 

Our study demonstrates a novel avenue for removing grain boundaries - simply turning on and off the self-propulsion of the active particles. Hence by simply controlling the energy source for the active particles, via e.g. the availability of fuel or an external light source, active particles can be used to grow large-scale single-domain crystals.

We acknowledge funding from the Dutch Sector Plan Physics and Chemistry, and funding from a NWO-Veni grant. We thank Vasileios Prymidis, Frank Smallenburg and Sela Samin for careful reading of the manuscript.

\end{document}